\documentclass[prd,aps,twocolumn,showpacs,floats,floatfix]{revtex4}
\usepackage[dvips]{graphicx}
\usepackage{amsmath}

\begin{document}
 
\title{Dark-Matter Admixed Neutron Stars}

\author{S.-C. Leung, M.-C. Chu, L.-M. Lin}
\affiliation{Department of Physics and Institute of Theoretical Physics, 
The Chinese University of Hong Kong, Hong Kong, China} 

\date{\today}

\begin{abstract} 
We study the hydrostatic equilibrium configuration of an admixture of 
degenerate dark matter and normal nuclear matter by using a 
general relativistic two-fluid formalism.  
We consider non-self-annihilating dark matter particles of mass $\sim 1$ GeV. 
The mass-radius relations and moments of inertia of these dark-matter admixed
neutron stars are investigated and the
stability of these stars is demonstrated by performing a radial 
perturbation analysis. 
We find a new class of compact stars which consists 
of a small normal matter core with radius of a few 
km embedded in a ten-kilometer-sized dark matter halo.  These stellar objects
may be observed as 
extraordinarily small neutron stars that are incompatible with realistic 
nuclear matter models.

\end{abstract}


\maketitle{}

%%%%%%%%%%%%%%%%%%%%%%%%%%%%%
%\section{Introduction}
%%%%%%%%%%%%%%%%%%%%%%%%%%%%%%%%%%%%%%%%%%%%%%%%%%%%% 
%\paragraph*{Introduction.} 
%%%%%%%%%%%%%%%%%%%%%%%%%%%%%%%%%%%%%%%%%%%%%%%%%%%%% 

{\it Introduction.}---By now the existence of dark matter (DM) 
has been well established, with a large amount of evidences such as 
galactic rotation curves, cosmological structure and gravitational lensing. 
However, the properties of DM including their mass and interactions 
are still unknown. It is thus of great interest to constrain the properties 
of DM through direct or indirect methods.

Direct methods search for the signals of DM-nucleus scatterings in
Earth-based detectors. The latest experimental results are not
conclusive. The data from the DAMA \cite{DAMA2008} and CoGeNT \cite{CoGeNT2011} 
experiments are consistent with detecting
light DM particles with mass $\sim 10$ GeV, 
which are incompatible with the null results from CDMS \cite{CDMS2011} and 
XENON \cite{XENON2010}. 
Nevertheless, it has recently been suggested that isospin-violating DM may be 
the key to reconciling the experimental results \cite{Feng2011,Frandsen2011}. 
On the other hand, indirect methods are based on the effects of DM on the 
properties of stellar objects such as the Sun. For example, the effects of 
low-mass ($\sim 5$ GeV) asymmetric DM particles on the solar composition, 
oscillations, and neutrino fluxes have been considered recently 
\cite{Frandsen2010,Cumberbatch2010,Taoso2010}. 

One indirect method that is gaining attention in recent years is to study the 
effects of DM on compact stars.
The effects due to different DM models have been considered. For example, 
self-annihilating DM inside compact stars can heat the stars, and hence affect 
the cooling properties of compact stars \cite{Kouvaris2008,Bertone2008}. 
On the other hand, non-self-annihilating DM, such as asymmetric DM 
\cite{Kouvaris2011} and mirror matter \cite{Ciarcelluti2011}, would simply 
accumulate inside the stars and affect the stellar structure. 
Constraints have been set by connecting the observed properties of compact 
stars with DM parameters.

It should be noted that neutron stars with a DM core are inherently two-fluid 
systems where the normal matter (NM) and DM couple essentially only through 
gravity. The technique used in recent studies of the structure of 
these dark-matter admixed neutron stars (DANS) is based on an {\it ad hoc} 
separation of the Tolman-Oppenheimer-Volkoff (TOV) equation into two different 
sets for the normal and dark components inside the star 
\cite{Lavallaz2010,Ciarcelluti2011}.
This approach is motivated by the similarity of the structure equations 
between the relativistic and Newtonian ones, but it is not derived from 
first principle. 
%%%Note that the derivation of the TOV equation is based on 
%%%the assumption that the star is described by a single fluid with a unique 
%%%four velocity, while NM and DM in general have different four 
%%%velocities. 
In fact, a general relativistic two-fluid formalism is available 
\cite{Carter1989} and has been employed in the study of superfluid neutron 
stars (e.g., \cite{Comer1999,Andersson2001}), where the two fluids are normal 
and superfluid nuclear matter.
This approach is not only more appealing from a 
theoretical point of view but also able to 
extend easily the study of dynamical properties 
of these stars in a self-consistent 
general relativistic framework. 
Here we study the structure and stability of DANS in general. 
Besides the scenario where a DM core exists inside a neutron star, we also 
study the scenario where NM is in the core of a DM
dominated compact star. 
It should be pointed out that the main focus of this paper is to study the 
equilibrium properties and observational signatures of these theoretical 
objects (see also \cite{Narain2006} for a study of compact stars made of fermionic DM). 
The formation process of these objects requires further investigation.

%%We shall investigate how the observed properties of 
%%compact stars could be used to distinguish the different scenarios. 

%%%%%%%%%%%%%%%%%%%%%%%%%%%%%%%%%%%%%
%\section{Formulation}
%%%%%%%%%%%%%%%%%%%%%%%%%%%%%%%%%%%%%%%%%%%%%%%%%%%%% 
%\paragraph*{Formulation.} 
%%%%%%%%%%%%%%%%%%%%%%%%%%%%%%%%%%%%%%%%%%%%%%%%%%%%%

{\it Formulation.}---To study a two-fluid compact star, we adopt the 
formulation given in \cite{Comer1999}, which was initially constructed to 
study general relativistic superfluid neutron stars. 
Here we shall briefly summarize the formalism and refer the reader to 
\cite{Comer1999} for more details. 
The central quantity of the two-fluid formalism is the master function 
$\Lambda(n^2 , p^2 , x^2)$, which is formed by three scalars, 
$n^2 = - n_{\alpha}n^{\alpha}$, $p^2 = - p_{\alpha}p^{\alpha}$, and 
$x^2 = - n_{\alpha}p^{\alpha}$. The four vectors $n^{\alpha}$ and $p^{\alpha}$ 
are the conserved NM and DM number density currents respectively. 
The master function is a two-fluid analog of the equation of state (EOS)
and $-\Lambda$ is taken to be the thermodynamic energy density. 

\begin{figure}
\centering
\includegraphics[width=7cm, height=5cm]{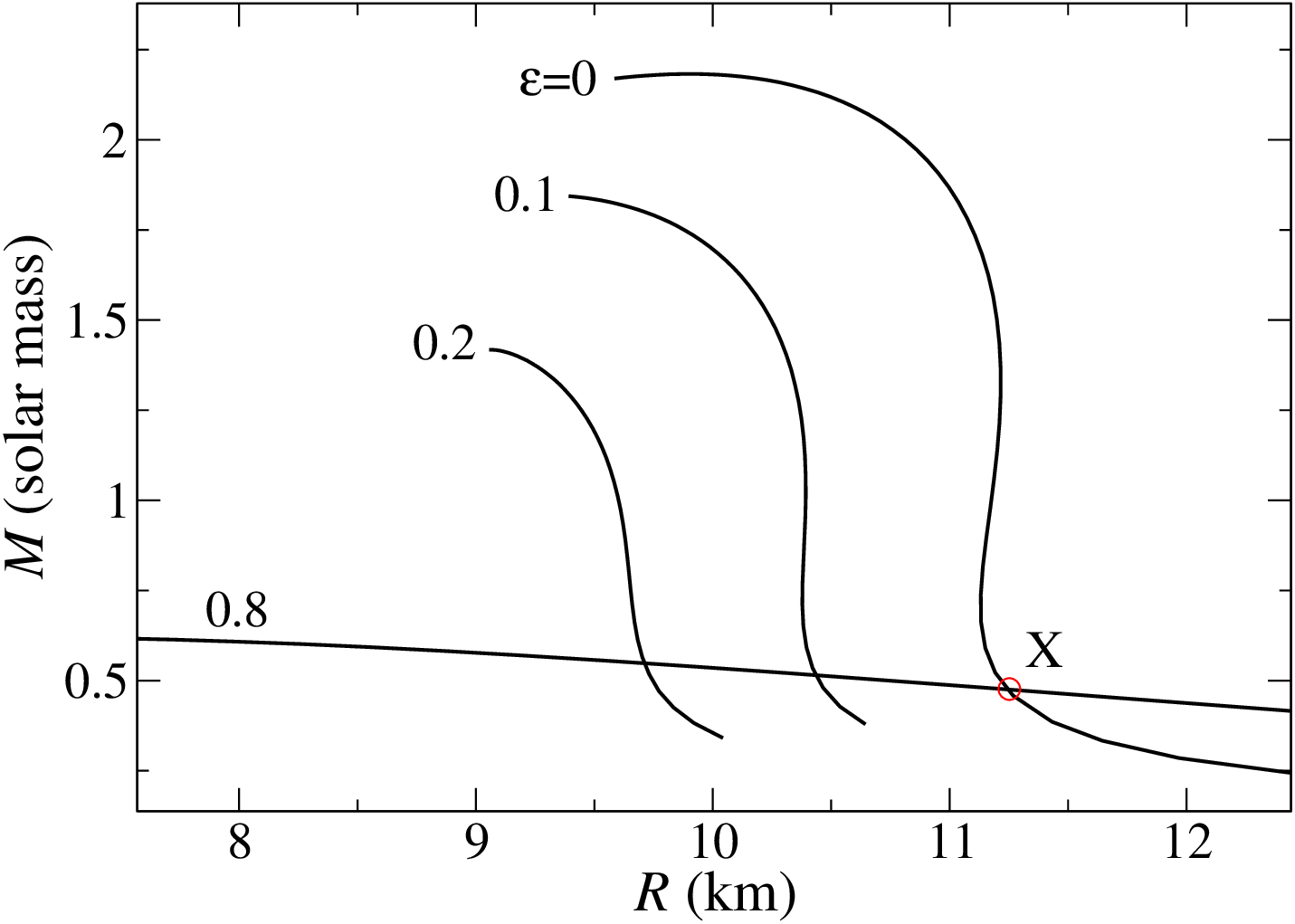}
\caption{Mass-radius relations for different amount of DM specified 
by the parameter $\epsilon$ (see text). The DM particle mass is $m_X=1$ GeV.
The density profiles for the ordinary neutron star ($\epsilon=0$) and 
DM dominated star ($\epsilon=0.8$) at the point X are shown in 
Fig.~\ref{fig:profile2}. }
\label{fig:MR_relation}
\end{figure}

%%%%%%%%%%%%%
% It should be noted that in the original work for superfluid neutron stars 
% \cite{Comer1999}, $n^{\alpha}$ stands for the superfluid-neutron number 
% density currents, while $p^{\alpha}$ is the number density current for a 
% conglomerate of all other charged constituents. 
% Here we only focus on the effects of a DM core on the structure of 
% neutron stars and do not take nuclear-matter superfluidity into account. 
% This should be a good approximation since superfluidity would have little
% effect on the global structure of neutron stars such as their masses and 
% radii. However, as we shall see, a DM core would change the static 
% properties of neutron stars significantly. 
%%%%%%%%%%%%%%%%%%%%

For a static and spherically symmetric spacetime $ds^2 = - e^{\nu(r)} dt^2 + 
e^{\lambda(r)}dr^2 + r^2 ( d\theta^2 + \sin^2\theta d\phi^2)$, the structure 
equations for a two-fluid compact star are given by \cite{Comer1999}
\begin{eqnarray}
A^0_0 p' + B^0_0 n' + \frac{1}{2} (B n + A p) \nu ' = 0,  
\nonumber\\
C^0_0 p' + A^0_0 n' + \frac{1}{2} (A n + C p) \nu ' = 0,  
\nonumber\\
\lambda^{'} = { {1 - e^\lambda} \over r } - 8\pi r e^\lambda \Lambda , 
\nonumber\\
\nu^{'} = - { {1 - e^\lambda} \over r } + 8\pi r e^\lambda \Psi , 
\label{eq:fluid_eq}
\end{eqnarray}
where the primes indicate derivative with respect to $r$, and the coefficients
$A$, $B$, $C$, $A^0_0$, $B^0_0$, and $C^0_0$ are functions of the master 
function. Their expressions are given by Eqs. (3) and (25) in 
\cite{Comer1999}. The generalized pressure $\Psi$ is computed by 
Eq. (18) in \cite{Comer1999}.

The EOS information $P=P(\rho)$ (with $P$ and $\rho$ being the pressure
and energy density respectively) needed in the standard relativistic-star 
calculation based on the TOV equation is now replaced by the master 
function $\Lambda(n^2 ,p^2, x^2)$. 
We assume no interaction between NM and DM except for gravitation. The master
function does not depend on the scalar $x^2=-n_{\alpha}p^{\alpha}$ and 
is separable in the sense that $\Lambda(n^2 , p^2) = \Lambda_{\rm NM}(n^2) 
+ \Lambda_{\rm DM}(p^2)$, $\Lambda_{\rm NM}(n^2)$ and 
$\Lambda_{\rm DM}(p^2)$ being the negative of energy densities of NM
and DM at a given number density respectively. 
We use the APR EOS \cite{Akmal1998} for NM
and assume that the DM component of the star is formed by non-self-annihilating DM
governed by an ideal Fermi gas. 
As discussed earlier, DM candidates in the mass range of a few GeV are of
great interest recently. We shall thus consider fermionic
DM particles of mass $m_X \sim 1$ GeV in this work.

%%%%%%%%%%%%%%%%%%%%%%%%%%%%%%%%%%%
%\section{Results}
%%%%%%%%%%%%%%%%%%%%%%%%%%%%%%%%%%%
%\paragraph*{Results.}
%%%%%%%%%%%%%%%%%%%%%%%%%%%%%%%%%%%%

{\it Results.}---In Fig. \ref{fig:MR_relation} we show the mass-radius 
relations of DANS for different amount of DM specified by the parameter 
$\epsilon = M_{\rm DM}/(M_{\rm NM} + M_{\rm DM})$. Here $M_{\rm NM}$ and 
$M_{\rm DM}$ are calculated by the product of the particle mass and total 
number of particles for NM and DM in the star. They may be 
referred to as the baryonic masses for NM and DM (though it should be noted
that DM is nonbaryonic). In the figure, $M$ is the gravitational mass and 
$R$ is the radius of the star. The DM particle mass $m_X=1$ GeV is fixed. 
The case $\epsilon=0$ corresponds to ordinary neutron star models 
constructed using the APR EOS without DM. We see that the existence of a DM 
core would lower the maximum stable mass allowed by this EOS. The DANS
would also have smaller radii. For the maximum stable mass configuration, 
$M$ and $R$ are decreased by about 35\% and 9\% respectively as
$\epsilon$ changes from 0 to 0.2. 
For the case $\epsilon=0.8$, the stars are DM dominated compact stars, with
NM concentrated in the core. 
It is seen clearly that the mass-radius relation of these stars is 
different from that of NM dominated stars qualitatively. 

\begin{figure}
\centering
\includegraphics[width=7cm, height=5cm]{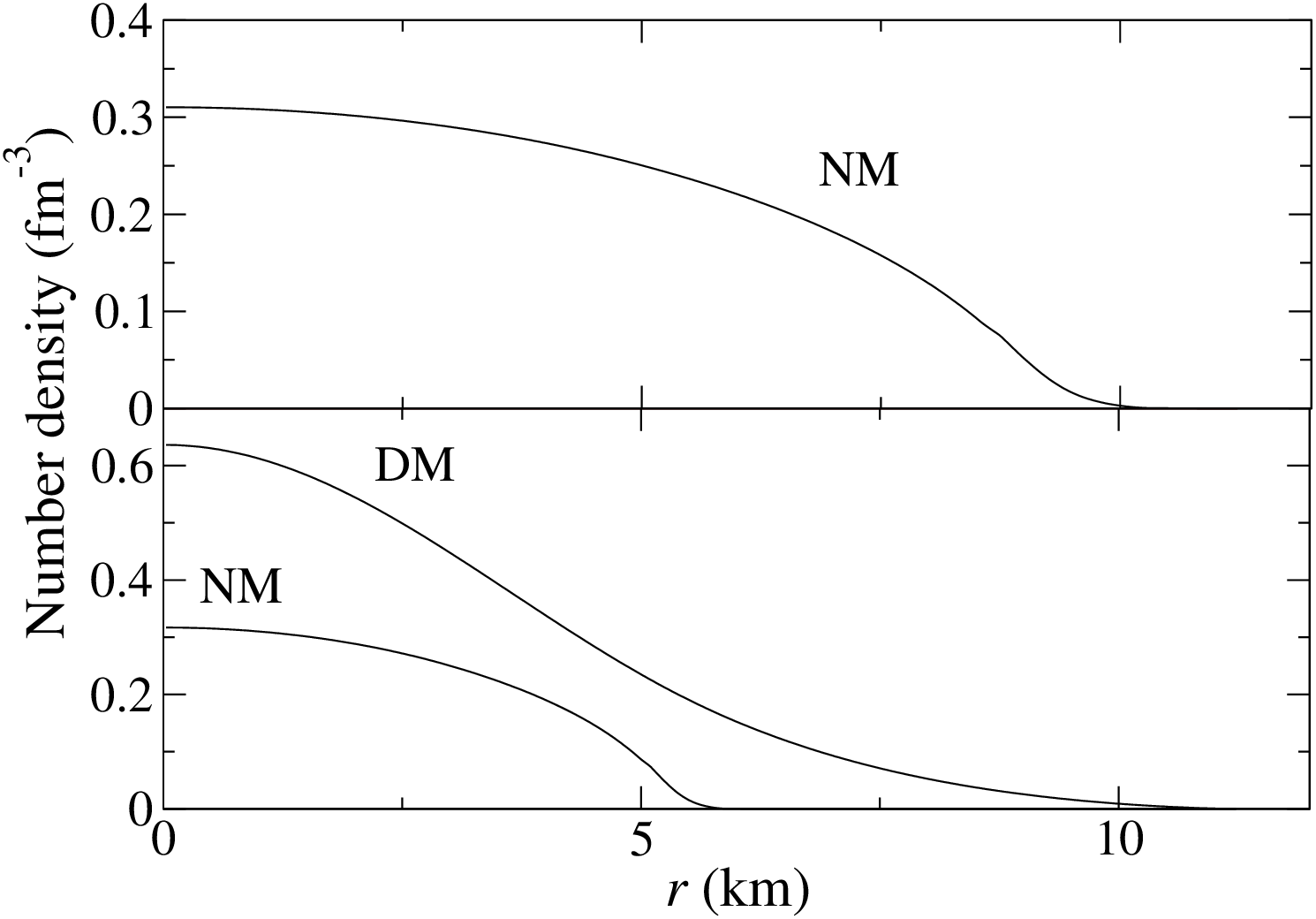}
\caption{Upper panel: Density profile for the ordinary neutron star 
($\epsilon=0$) at the point X in Fig.~\ref{fig:MR_relation}. 
Lower panel: NM and DM density profiles for the DM dominated star 
($\epsilon=0.8$) at the same point.} 
\label{fig:profile2}
\end{figure}

In Fig.~\ref{fig:MR_relation}, the circle marked by X is the intersection 
point between the curves $\epsilon=0$ and 0.8. 
While the ordinary neutron star ($\epsilon=0$) and dark-matter dominated star 
($\epsilon=0.8$) at X have the same $M$ and $R$, their internal 
structures in fact differ significantly. 
Fig.~\ref{fig:profile2} shows the density profiles of these two stellar
configurations, the upper (lower) panel 
corresponding to the ordinary neutron star (DANS)
model. For the case $\epsilon=0.8$ (lower panel), it is seen 
that a small NM core is embedded in a ten-kilometer sized DM halo. 
Since $M$ and $R$ of the two stars are the same, it would seem impossible 
to distinguish them based on their gravitational effects on other nearby
stellar objects. 
However, the visible radius of the DM dominated star (defined by the radius of 
the NM core) is $0.56 R$ and the total mass enclosed in the NM core is 
$0.72 M$. The two stars can be distinguished by measuring the 
gravitational redshift of spectral lines, since that produced near the 
surface of the NM core will be about $30\%$ 
larger than that of ordinary neutron stars.

\begin{figure}
\centering
\includegraphics[width=7cm, height=5cm]{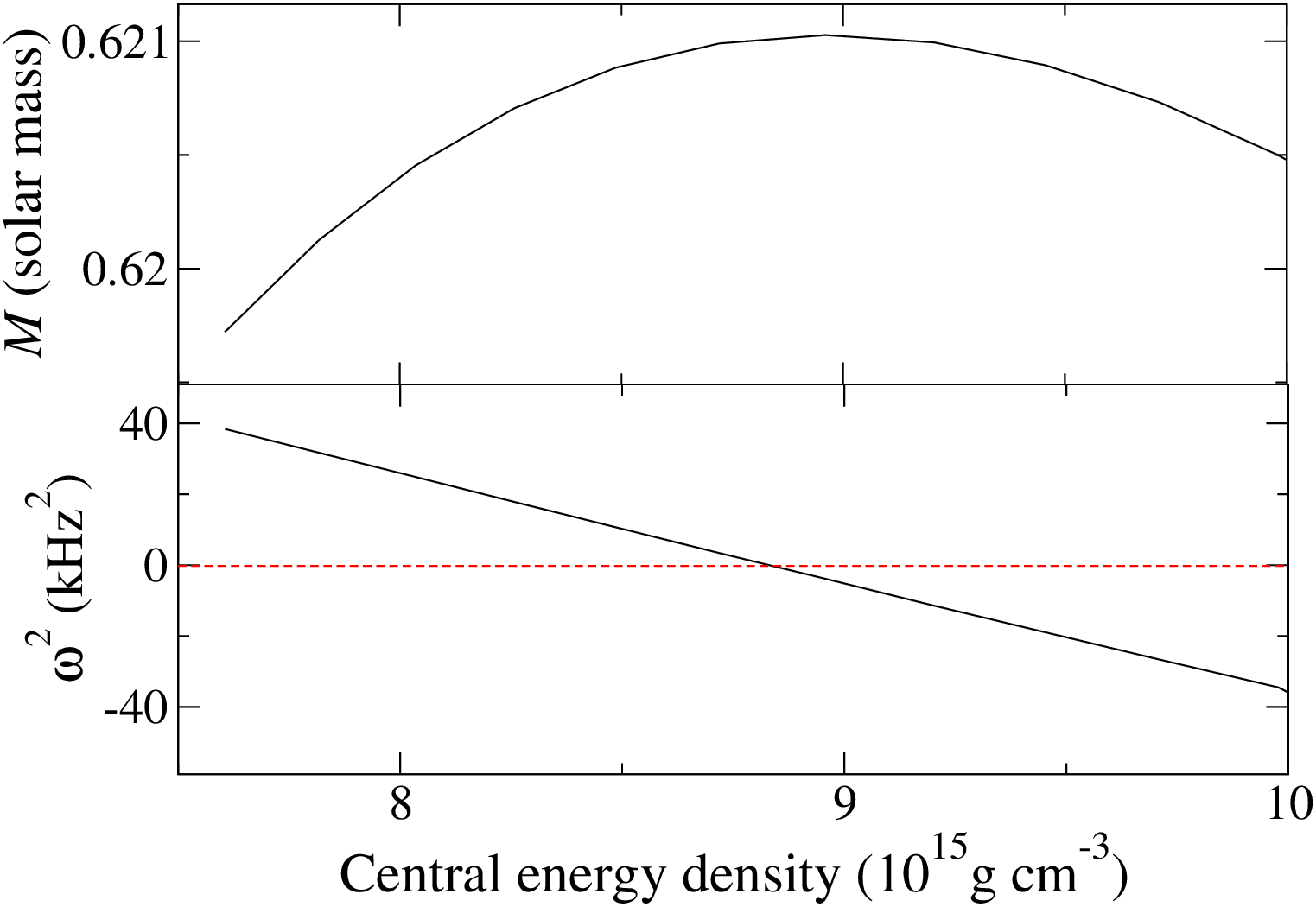}
\caption{Gravitational mass $M$ (upper panel) and squared frequency
of the fundamental mode $\omega_0^2$ (lower panel) are plotted against
the central energy density for the DM dominated sequence ($\epsilon=0.8$)
in Fig.~\ref{fig:MR_relation}. }
\label{fig:eigenvalue_M}
\end{figure}

In order to check the stability of these stars, we have solved the set of 
equations for radial perturbations of a two-fluid compact stars developed in 
\cite{Comer1999}. Similar to the one-fluid case, the 
problem is to solve for the eigenvalues $\omega^2$, where $\omega$ is the 
oscillation frequency of the star. We will present the details of our 
calculations and analysis elsewhere. 
Here we show our main results in Fig.~\ref{fig:eigenvalue_M}, where
$M$ is plotted against the central energy density in the upper panel
for the DM dominated sequence ($\epsilon=0.8$) in Fig.~\ref{fig:MR_relation}.
The lower panel plots the squared frequency of the fundamental mode 
$\omega_0^2$ for the same sequence. Similar to the one-fluid study for the 
stability of ordinary neutron stars, the point $\omega_0^2=0$ marks the onset 
of instability. Fig.~\ref{fig:eigenvalue_M} shows that $\omega_0^2$ passes 
through zero at the central density corresponding to the maximum mass 
configuration. Beyond this critical central density, the stars are unstable 
against radial perturbations. For lower central densities, such as the DM 
dominated star at the point X in Fig.~\ref{fig:MR_relation}, the stellar
configurations are all stable.

Besides the gravitational mass and radius, it is also interesting to consider
the moment of inertia $I$ of DANS since it is measurable and plays 
an important role in the physics of neutron stars. 
Bejger and Haensel \cite{Bejger2002} discovered an (approximately) 
EOS-independent formula relating $I$, $M$ and $R$. 
For ordinary neutron stars, they found that 
\begin{eqnarray}
\tilde{I} = \Bigg\{\begin{array}{ll} z/(0.1+2z) ~~& \textrm{if $z\le 0.1$,}\\
{2}(1+5 z)/{9} & \textrm{if $z>0.1$,}
\end{array}
\label{eq:BH_inertia}
\end{eqnarray}
where the scaled moment of inertia $\tilde{I} = I/MR^2$ and 
$z=(M/M_{\odot})({\rm km}/R)$. 
This universal formula was obtained by fitting a large set of realistic 
EOS models for nuclear matter, including the APR EOS used in this work. 
The moment of inertia of a rotating star in general relativity is commonly 
defined by $I = J/\Omega$, where $J$ and $\Omega$ are
the angular momentum and angular velocity respectively. In the slow rotation 
limit, $J$ scales with $\Omega$ linearly and $I$ is determined only by the 
nonrotating background quantities \cite{Hartle1967}.

\begin{figure}
\centering
\includegraphics[width=7cm, height=5cm]{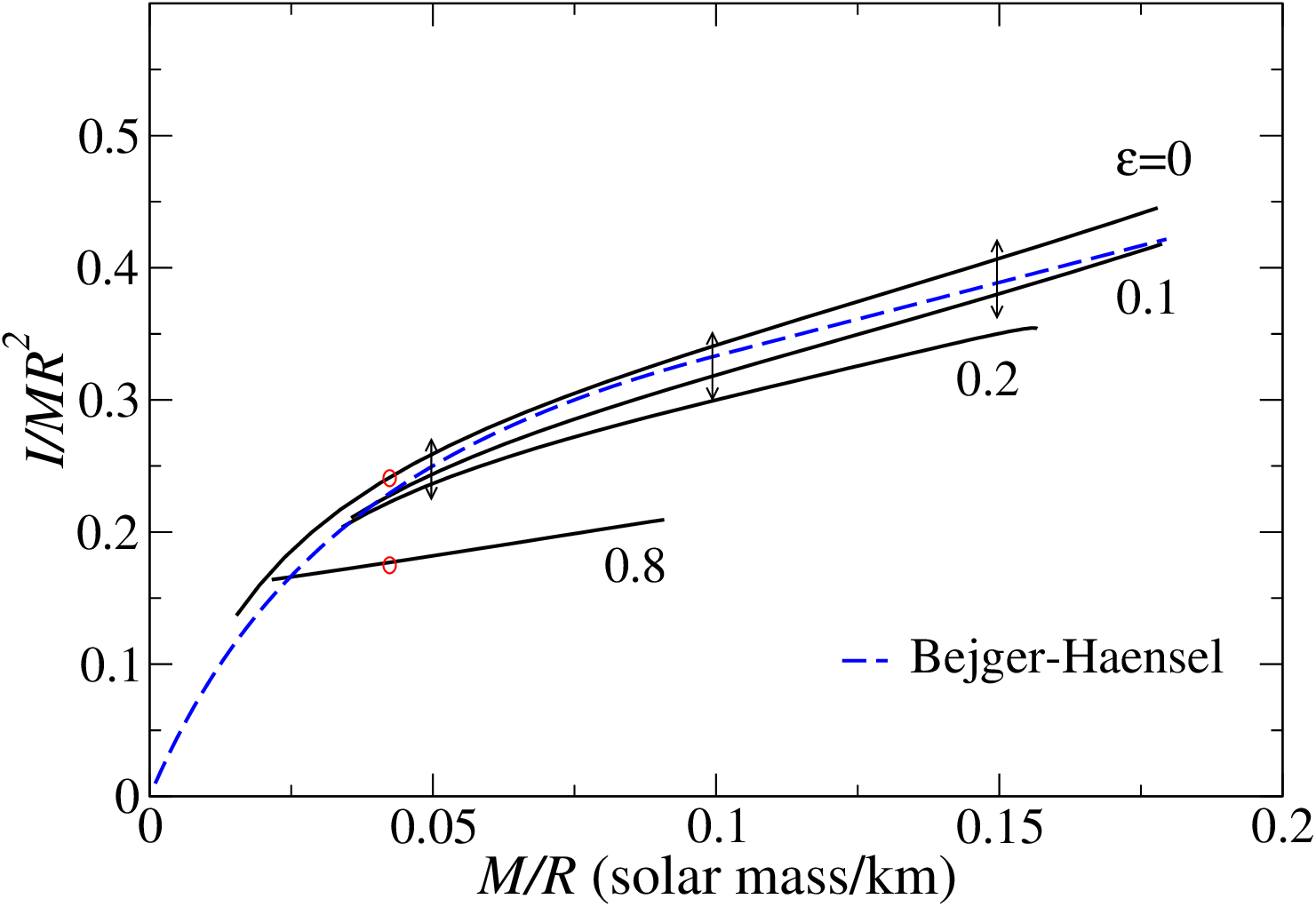}
\caption{Scaled moment of inertia $I/MR^2$ is plotted against
compactness $M/R$ for the DANS sequences shown in Fig.~\ref{fig:MR_relation}.
Each sequence is labeled by the parameter $\epsilon$.
The dashed line represents Eq.~(\ref{eq:BH_inertia}). The circles correspond
to the two stellar models ($\epsilon=0$ and 0.8) at the point X in 
Fig.~\ref{fig:MR_relation}.}
\label{fig:inertia_compact_APR}
\end{figure}

In the two-fluid case, individual NM and DM angular momenta ($J_{\rm NM}$ and 
$J_{\rm DM}$) can be defined \cite{Andersson2001}. 
In the slow rotation limit, the moments of inertia can also be defined by 
$I_{\rm NM}=J_{\rm NM}/\Omega_{\rm NM}$ and 
$I_{\rm DM}=J_{\rm DM}/\Omega_{\rm DM}$, where $\Omega_{\rm NM}$ and 
$\Omega_{\rm DM}$ are, respectively, the angular velocity of NM and DM.
$I_{\rm NM}$ and $I_{\rm DM}$ depend on the nonrotating background 
quantities of NM and DM separately. However, their definitions are 
meaningful only when the two fluids are non-interacting (as we assume
in this work). In the general situation, the angular momentum of each fluid 
would contain a contribution which depends on the coefficient $A$ in 
Eq.~(\ref{eq:fluid_eq}) and the relative velocity 
$\Omega_{\rm NM}-\Omega_{\rm DM}$ \cite{Andersson2001}. 
The coefficient $A$ vanishes only when the master function $\Lambda$ is 
independent of the scalar product $x^2 = -n_{\alpha}p^{\alpha}$.

With the moments of inertia of NM and DM defined individually as above, 
we can further define the total moment of inertia of DANS by 
$I=I_{\rm NM}+I_{\rm DM}$.   
In Fig.~\ref{fig:inertia_compact_APR}, we plot the scaled moment of inertia 
$\tilde{I}$ against compactness $M/R$ for the DANS sequences shown in 
Fig.~\ref{fig:MR_relation}. 
As in Fig.~\ref{fig:MR_relation}, the solid lines are sequences for different
amount of DM specified by $\epsilon$. The case $\epsilon=0$ corresponds to 
ordinary neutron stars. The dashed line corresponds to 
Eq.~(\ref{eq:BH_inertia}). 
The three vertical lines (with arrows) at $M/R=0.05$, 0.1 and 0.15 represent 
the range of values of $\tilde{I}$ obtained by the large set of EOS models 
which were used to obtain Eq.~(\ref{eq:BH_inertia}). They can be regarded 
as the error bars of Eq.~(\ref{eq:BH_inertia}) at those values of $M/R$.
The circles in the figure correspond to the ordinary neutron star 
($\epsilon=0$) and DM dominated star ($\epsilon=0.8$) at the point X in 
Fig.~\ref{fig:MR_relation}. 
While the scaled moment of inertia of ordinary neutron stars can be modeled 
approximately by Eq.~(\ref{eq:BH_inertia}), Fig.~\ref{fig:inertia_compact_APR} shows that 
$\tilde{I}$ of DANS depends sensitively on the amount of DM. 
In particular, for the DM dominated sequence $\epsilon=0.8$, the 
value of $\tilde{I}$ is significantly smaller than that allowed for 
ordinary neutron stars with the same compactness. This might lead to 
observational signatures of DM dominated compact stars. 
Further work is required to investigate the 
observational implications of DANS in details.

\begin{figure}
\centering
\includegraphics[width=7cm, height=5cm]{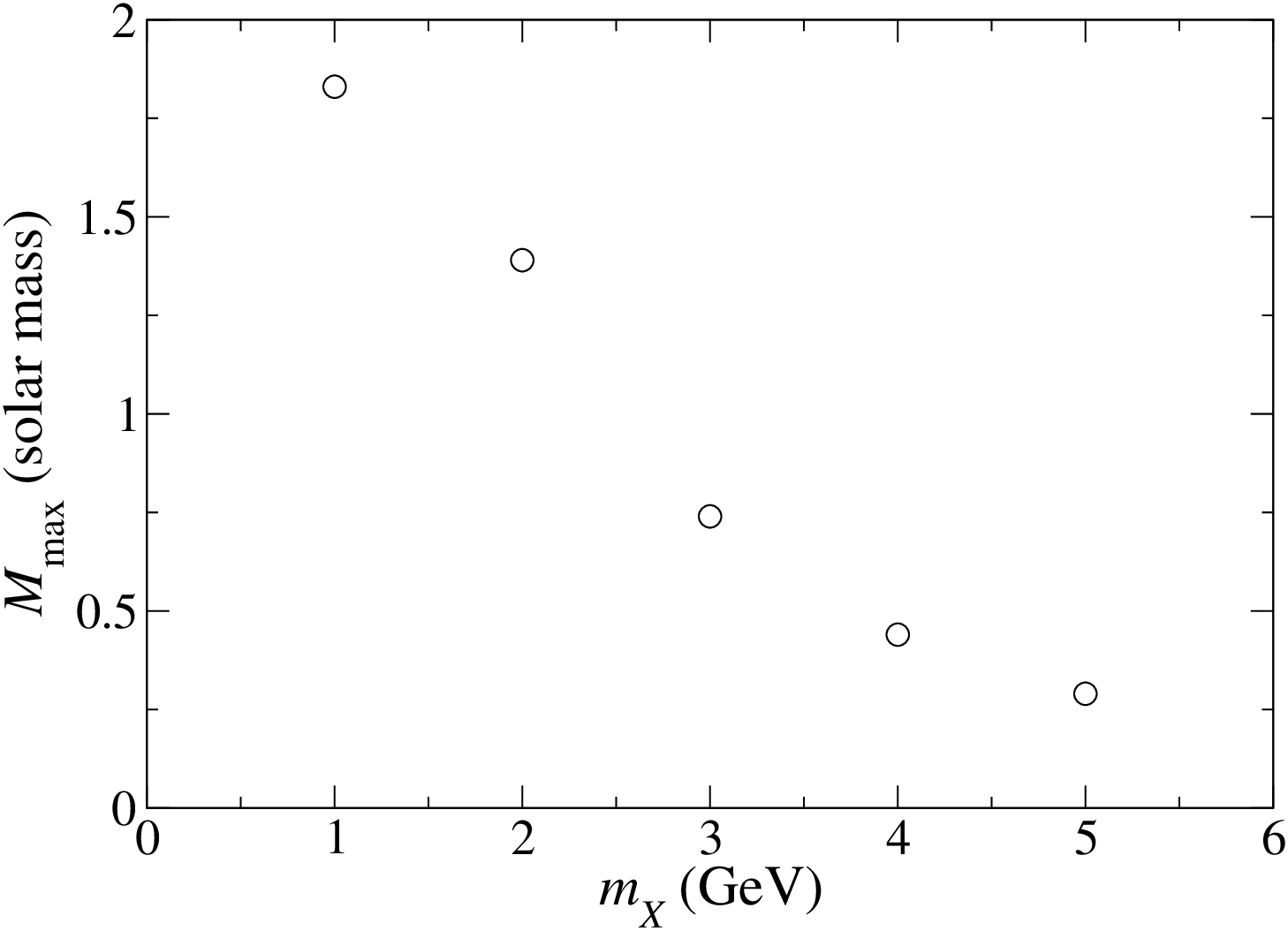}
\caption{Maximum stable mass $M_{\rm max}$ is plotted against the DM particle 
mass $m_X$ for a fixed amount of DM specified by $\epsilon=0.1$. } 
\label{fig:Mmax_eps0.1}
\end{figure}

So far we have focused our study on DM particle mass $m_X=1$ GeV. In view of 
the recent interest in DM candidates in the mass range of a few GeV, it is 
interesting to see how different DM particle masses in this range affect our 
results. In Fig.~\ref{fig:Mmax_eps0.1}, we plot the maximum stable mass 
$M_{\rm max}$ along a sequence of compact stars with $\epsilon=0.1$ as a 
function of $m_X$. 
Note that the sequence for $m_X=1$ GeV is shown in Fig.~\ref{fig:MR_relation}. 
For a given proportion of DM inside the stars, Fig.~\ref{fig:Mmax_eps0.1}
shows that a higher DM particle mass in general leads to a smaller maximum 
stable mass. 
This can easily be understood by noting that, since the NM and DM are assumed
to be non-interacting (except through gravity), the DM core is supported only 
by its own degenerate pressure. 
It is well known that the maximum mass limit for a self-gravitating
Fermi gas decreases as the particle mass increases.
Hence, the onset of the collapse of a degenerate DM core is responsible for 
the dependence of $M_{\rm max}$ on $m_X$ as seen in 
Fig.~\ref{fig:Mmax_eps0.1}. 
It should also be noted that, while the pressure of NM within the DM core 
does not contribute to supporting the weight of the DM core, the mass of 
the NM fluid does enhance the collapse of the DM core. 

%%%%%%%%%%%%%%%%%%%%%%%%%%%%%%%%%%%%%%%%%%%%%%%%%%%%%%%%%%%%%%%%%%%
%\section{Conclusions}
%%%%%%%%%%%%%%%%%%%%%%%%%%%%%%%%%%%%%%%%%%%%%%%%%%%%%%%%%%%%%%%%%%%

{\it Discussion and Conclusions.}---In summary, we have studied the effects of a 
degenerate DM core formed by non-self-annihilating DM particles of mass $\sim 1$ GeV
upon the structure of neutron stars. 
The structure of these DANS depends strongly on the size of the DM core. 
In particular, we found a new class of compact stars which are DM dominated ---
a NM core embedded in a ten-kilometer sized DM halo. 
The stability of these stars has been checked by performing a radial
perturbation analysis. 
These DM dominated stars have rather different mass-radius relations and 
(scaled) moments of inertia comparing to ordinary neutron stars. 
A distinctive property of these stars is their small NM core radius of
about a few km, from which thermal radiation could be observed. 
The detection of a compact star with a thermally radiating surface of such 
a small size could provide a strong evidence for their existence. 

Could DANS be formed in the first place? 
To answer this question, one needs to 
consider the effects of DM on the 
stellar formation process. In fact, 
the heating effects due to DM 
annihilation on stellar formation have been studied in recent years 
\cite{Spolyar2008}.
The result is the prediction of a new phase of stellar evolution during which
a protostar is supported by DM heating. What if one replaces the annihilating
DM model in \cite{Spolyar2008} by non-self-annihilating DM? 
How would such a DM core affect the stellar evolution? 
Would the DM core survive the supernova explosion of massive stars and 
form DANS as studied in this paper? 
These are challenging questions that deserve further investigation.

%%%%%%%%%%%%%%%%%%%%%%%%%%%%
%\section{Acknowledgement}
%%%%%%%%%%%%%%%%%%%%%%%%%%%

{\it Acknowledgement.}---This work is partially supported by 
a grant from the Research Grant Council of the Hong Kong Special 
Administrative Region, China (Project No. 400910).

%\newpage

%%%%%%%%%%%%%%%%%%%%%%%%%%%%%%%%%%%%%% 
%% reference 
%%%%%%%%%%%%%%%%%%%%%%%%%%%%%%%%%%%%%% 
\bibliographystyle{prsty}

\end{document}